\documentstyle[preprint,eqsecnum,aps,epsf]{revtex}
\input{psfig}
\tighten                              
\newcommand{\beq}{\begin{equation}}
\newcommand{\eeq}[1]{\label{#1} \end{equation}}
\newcommand{\beqar}{\begin{eqnarray}}
\newcommand{\eeqar}[1]{\label{#1} \end{eqnarray}}
\begin{document}
\draft         
\preprint{\vbox{\hbox{IFT--P.077/97}}}
%
%
\title{ Testing the Resolving Power of 2-D $K^+ K^+$ Interferometry$^*$
 \footnotetext{ *Partially supported by Conselho Nacional de Desenvolvimento
Cient\'\i fico e Tecnol\'ogico (CNPq), and Funda\c c\~ao de Amparo \`a 
Pesquisa do Estado de S\~ao Paulo (FAPESP), Brazil}\\[5ex] }

\author{ Sandra.\ S.\ Padula and Cristiane G. Rold\~ao} 
\address{Instituto de F\'{\i}sica Te\'orica, 
Universidade  Estadual Paulista, \\  
Rua Pamplona 145, CEP 01405-900 S\~ao Paulo, Brazil.}
\date{\today}
\maketitle
\begin{abstract}
Adopting a procedure previously proposed to quantitatively study 
two-dimensional pion interferometry \cite{pgplb95}, an equivalent 
2-D $\chi^2$ analysis was performed  to test the resolving power 
of that method when applied to less favorable conditions, i.e., 
if no significant contribution from long lived resonances 
is expected, as in kaon interferometry. 
For that purpose, use is made of the preliminary E859 
$K^+ K^+$ interferometry data \cite{vince} from $Si+Au$ 
collisions at 14.6 AGeV/c. 
As expected, less sensitivity is achieved in the present case, 
although it still is possible to distinguish 
two distinct decoupling geometries. The present analysis seems to favor 
scenarios with no resonance formation at the AGS energy range, 
if the preliminary $K^+ K^+$ data are confirmed. 
The possible compatibility of data with zero decoupling 
proper time interval, conjectured by the 3-D experimental analysis 
\cite{vince}, 
is also investigated and is ruled out when considering more realistic 
dynamical models with expanding sources. 
These results, however, clearly evidence the important influence of 
the time emission interval on the source effective transverse dimensions. 
Furthermore, they strongly emphasize that the static 
Gaussian parameterization, commonly used to fit data, cannot be trusted 
under more realistic conditions, leading to 
distorted or even wrong interpretation of the source parameters! 

\end{abstract}
\narrowtext
     
 
\section{Introduction}

     The second-order interferometry of identical particles 
is a powerful tool for probing 
the space-time zone from which they were emitted \cite{zajc:boal}. 
Almost two decades ago, 
it was suggested as a possible signature of the Quark Gluon Plasma 
(QGP), expected to be formed in high energy Nucleus-Nucleus collisions,
by probing the expected large space-time dimensions of the emitting system 
at freeze-out. About ten years ago, when the first $O+Au$ runs 
from CERN/SPS became available, there were  
expectations that we could be seeing its formation, 
particularly from $\pi \pi$ interferometry\cite{na35OAu}. 
However, due to limited statistics, the correlation function 
at that time had to be 
projected in  one dimension only, leading to ambiguity in describing the 
overall behavior of interferometry data, i.e., they could be equally well 
described by two very distinct freeze-out scenarios \cite{gyupa89}. 
One of them reflected a dynamical model in which 
the pions were formed after the hadronization of the QGP and the other 
one considered, instead,the formation of a hadronic gas of 
resonances. 

On the other hand, several 
studies\cite{pg:nioc,hamapad,pratthpb} have shown that dynamical models 
considering expanding systems, would lead to effects that could 
dramatically distort the  two-particle correlation function. Among them, 
the most significant effect \cite{pg:nioc} was caused by long 
lived resonances, which later decayed into 
the observed particles. As a side consequence
of this study, it was suggested to use pion interferometry to probe 
resonance formation at energies where their fractions were 
unknown\cite{pgqm90}. Once again ambiguity in separating different 
scenarios emerged, evidencing symptoms of urgency for very accurate and 
high statistics data, which has become available more recently, 
allowing for multi-dimensional analyses. Nevertheless, together with improved 
data, more precise theoretical and phenomenological tests were required, 
leading to the method suggested in Ref. \cite{pgplb95}, in which 
a two-dimensional $\chi^2$ analysis was proposed to study the resolving 
power of pion interferometry. 
For that purpose, two dynamical scenarios were considered which   
predicted similar correlation functions, 
although the underlying decoupling geometries differed considerably. 
In one, long lived resonances were neglected, while in the other,
a resonance gas with fractions predicted by 
the Lund model\cite{lund} was considered. 

Nonetheless, to quantify the differences in terms of a $\chi^2$ 
interferometric analysis, the contribution of long lived resonances decaying 
into pions seemed to be essential. This fact led to 
the question whether the resolving power of the method would remain high under 
less favorable conditions, i.e., if only shorter lived resonances would 
contribute to the particle yield, as is the case of $K^+ K^+$ interferometry, 
This is precisely the goal of this paper.
Furthermore, the method is applied to test the hypothesis of 
zero time emission interval, 
suggested by the experimental fit using 3-D static Gaussian 
parameterization \cite{vince}. 
In this study the influence of the time emission interval on the 
transverse radius parameter emerged naturally and 
another very important point was clearly emphasized, i.e., 
the static Gaussian 
parameterization, popularly used to fit data, is usually misleading 
in more realistic situations, and results in distorted or even wrong 
interpretation of the source parameters! 
Prior to reach these points, 
however, we present a brief 
summary of the theoretical model underlying the analysis, the so-called 
Covariant Current Ensemble Formalism\cite{pg:nioc,ccef} and a brief review of 
the method discussed in Ref.\cite{pgplb95}. This is then adapted 
to the present case, in which use 
is made of the preliminary E859 bidimensional data on 
$K^+ K^+$ interferometry from the AGS/BNL.

\section{The Covariant Current Ensemble Formalism} 

 Under idealized conditions the  correlation function, 
$C_2({\boldmath k}_1,{\boldmath k}_2)$, of two identical 
bosons probes  their decoupling or freeze-out space-time
distribution, $\rho(x)$, through
$C_2({\boldmath k}_1,{\boldmath k}_2)= 1 + | \rho(k_1 - k_2) |^2$.
However, in actual high energy reactions, final state interactions,
correlations between coordinate and momentum variables, 
and resonance production
distort this ideal interference pattern, corresponding only to 
Bose-Einstein symmetry(see e.g. \cite{gyupa89}-\cite{pratthpb}). 
This may lead to erroneous interpretation about the underlying 
information on the decoupling geometry coming from the second-order 
interference pattern. 
In realistic cases, then, it is mandatory to employ more 
general formalisms\cite{gyupa89,pg:nioc,hamapad,pratthpb,ccef}, flexible 
enough to include such non-ideal effects, reflecting model dependent 
scenarios. In the Covariant Current Ensemble formalism,  
the  correlation function can be  expressed as\cite{pg:nioc,ccef}  
\begin{equation}
C(k_1,k_2) = \Upsilon(q)\left(
1 + \frac{|G(k_1,k_2)|^2}{G(k_1,k_1) G(k_2,k_2)} \right) 
\; \; , \label{cce}\end{equation} 
where $\Upsilon(q)  = (q_c/q) / ( e^{q_c/q} -1)$ is the  
Gamow factor that distorts the interference pattern due to final 
state Coulomb interactions, with $q_c = 2\pi 
\alpha m$ and $q=(-(k_1-k_2)^2)^{1/2}$. 

In general, when resonances are produced, 
the  complex amplitude, $G(k_1,k_2)$, can be written as
\begin{equation}
G(k_1,k_2) \approx \langle  \sum_{r} f(K^+/r)
 \left(1-iq u_r / \Gamma_r \right)^{-1}
e^{i q x_r} j^*_0(u^\mu_f k_{1 \mu}) j_0(u^\mu_f k_{2 \mu}) \rangle
\; \; , \label{ceresgij}\end{equation}
where $f(K^+/r)$ is the fraction of the observed $K^+$'s arising from
the decay of a resonance of type $r$, which freezes-out with final four 
velocity $u_r^\mu$. It should be noted that, in the absence of resonances, 
the sum in eq.(\ref{ceresgij}) reduces to only one term, $f(K^+)=1$.
The currents, $j_0(u_f k_i)$, contain information about the 
production dynamics. 

The ensemble average in the above notation is performed 
by using the following parameterization for the 
implicit break-up distribution\cite{gyupa89,pg:nioc}

\begin{equation}
D(x,p) \propto \;
 \exp\left\{-\frac{\tau^2}{ \Delta \tau^2}-\frac{(y-y^*)^2}{2 Y_c^2} 
- \frac{(\eta - y)^2}{2 \Delta \eta^2}
- \frac{x_T^2}{R_T^2}\right\}  \delta(E - E_{\boldmath p}) 
\delta^2({\boldmath p}_T) 
\; \;, \label{niioc}\end{equation}
where $\tau =(t^2-z^2)^\frac{1}{2} $
is the freeze-out proper time, and
$\eta = \frac{1}{2} \log((t+z)/(t-z)), y = \frac{1}{2}\log((E+p_z)/(E-p_z))$ 
are the space-time and momentum rapidity variables, respectively. 
The correlation between these rapidities is estimated 
from the Lund model
to be  $\Delta \eta \approx 0.8$ \cite{pg:nioc}, 
$Y_c=0.7$,and $y^*_{cm}=0$. As regarding resonance fractions, the  Lund 
model\cite{lund} in the AGS range suggests essentially two 
contributions for that scenario, i.e., 
that $f(K^+_{direct})=0.5$,
$f(K^+/K^*)=0.5$. 

We recall that the transverse momentum in the 
more general model proposed in Ref.\cite{pg:nioc} is assumed to arise entirely
from  the finite momentum spread $\Delta p$ of the pion 
wave-packets. It should also be clarified that this model 
coincides with the Covariant Current Ensemble formalism  
in the case of minimum packets, when associating the momentum spread to the 
so-called pseudo-temperature, $T_{PS}$, 
through $\Delta p^2/m = T_{PS}$. This pseudo-thermal ansatz, however, was 
previously used in order to derive an analytical 
form for the correlation function\cite{ccef}. 
In the present analysis we are basically considering the Covariant 
Current Ensemble formalism but, since numerical calculations are 
carried out from the start, we consider the full thermal ansatz
instead, 
in which $T$ is the effective inverse transverse mass slope 
from the experimental fit, i.e., 
$T=0.18$ GeV \cite{vince:priv}, 
corresponding to an average momentum $\langle k_T \rangle
\approx 0.49$ GeV/c. It should be added that no clear difference could be 
seen when comparing the correlation functions corresponding to the 
thermal versus pseudo-thermal cases, in the same kinematical region. 
The currents in the thermal model may be written covariantly as
$
j_0(k) = \sqrt{u^\mu k_{\mu}} e^{-\frac{u^\mu k_{\mu}}{2mT}}
$. 

	By carring out the ensemble average in Eq. (\ref{ceresgij}) with the 
aid of (\ref{niioc}) and of $j_0(k)$ defined above, 
we obtain the expressions for $G(k_1,k_2)$ 
used in the numerical calculations: 
\begin{eqnarray}
G(k_1,k_2) &\propto& e^{-{\boldmath  q^2_T} R_T^2/4}
\int^{\infty}_{0} \tau \; d\tau \; e^{-\frac{\tau^2}{ \Delta \tau^2} }
\int^{+\infty}_{-\infty} dy \; e^{-\frac{(y-y^*)^2}{2 Y_c^2}} 
\int^{+\infty}_{-\infty} d\eta \; 
e^{- \frac{(\eta - y)^2}{2 \Delta \eta^2}} 
e^{i\tau(q_0 cosh\eta-q_L sinh\eta)} \nonumber \\
&& \sum_{r} f(K^+/r) \left(1-iq u_r/ \Gamma_r \right)^{-1}
\sqrt{[m_{1T}cosh(y_r-y_1)] [m_{2T}cosh(y_r-y_2]} 
\nonumber \\
&& \exp\{-m_{1T}cosh(y_r-y_1)/2T\} \; \exp\{-m_{2T} cosh(y_r-y_2)/2T\}
\; \; . \label{pp2}\end{eqnarray}

The single inclusive kaon distribution in this notation is 
$
P_1(k_i) \propto G(k_i,k_i) 
$, 
which can be written, with the help of eq. (\ref{niioc}), as 

\begin{eqnarray}
G(k_i,k_i) &\propto& \int^{\infty}_{0} \tau \; d\tau
e^{-\frac{\tau^2}{ \Delta \tau^2} }
\int^{+\infty}_{-\infty} dy  \; e^{-\frac{(y-y^*)^2}{2 Y_c^2}}
\int^{+\infty}_{-\infty} d\eta  \; 
e^{- \frac{(\eta - y)^2}{2 \Delta \eta^2}}
\nonumber \\
&& \sum_{r} f(K^+/r)  [m_{iT}cosh(y_r-y_i)] \;
 \exp\{-m_{iT}cosh(y_r-y_i)/T\} 
\; \;. \label{p1}\end{eqnarray}

The aim of the present study is to test if multidimensional kaon 
interferometry can discriminate scenarios including resonances 
from those in which they are absent, even 
in the much less striking limit of no significant long lived resonance 
contribution to the kaon yield.
For doing this, we apply the method suggested in Ref.\cite{pgplb95} 
to extract the rms transverse radius, $R_T$, at decoupling and the rms 
decoupling proper time interval, $\Delta \tau$. 
Note that we assume  implicitly that the chaoticity parameter
$\lambda=1$  throughout our analysis.

\section{$\chi^2$ Analysis}

To compare theoretical correlation functions with data
projected onto two of the six dimensions, we must compute the 
projected correlation function as 

\begin{equation}
C_{proj}(q_T,q_L) = \frac{ \int d^3 k_1 d^3 k_2 
P_2({\boldmath k}_1,{\boldmath k}_2) 
\;A_2(q_T,q_L;{\boldmath k}_1,{\boldmath k}_2)}
{\int d^3 k_1 d^3 k_2 
P_1({\boldmath k}_1)P_1({\boldmath k}_2) 
\;A_2(q_T,q_L;{\boldmath k}_1,{\boldmath k}_2)}
\; \; , \label{c35}\end{equation}
where $P_1$ and $P_2$ are the one and two kaon inclusive distributions,
and $A_2$ is the experimental two kaon binning and acceptance function.
All calculation were performed using the Monte Carlo importance sampling
method adopted in the CERES code\cite{pg:nioc}.

The acceptance function for the E859 experiment was approximated\cite{vince} 
by 
\begin{equation}
A_2(q_T,q_L;{\boldmath k}_1,{\boldmath k}_2) = 
A_1({\boldmath k}_1) A_1({\boldmath k}_2) 
\Theta(22-|\phi_1-\phi_2|)
 \delta(q_L-|k_{z1}-k_{z2}|)  
  \delta(q_T-|{\boldmath k}_{T_1}-{\boldmath k}_{T_2}|)  
\; \; .\label{a2exp}\end{equation}

\vskip .7cm
\begin{figure}[htb]
\protect
\epsfxsize=0.8\textwidth
\begin{center}
\leavevmode
\epsfbox{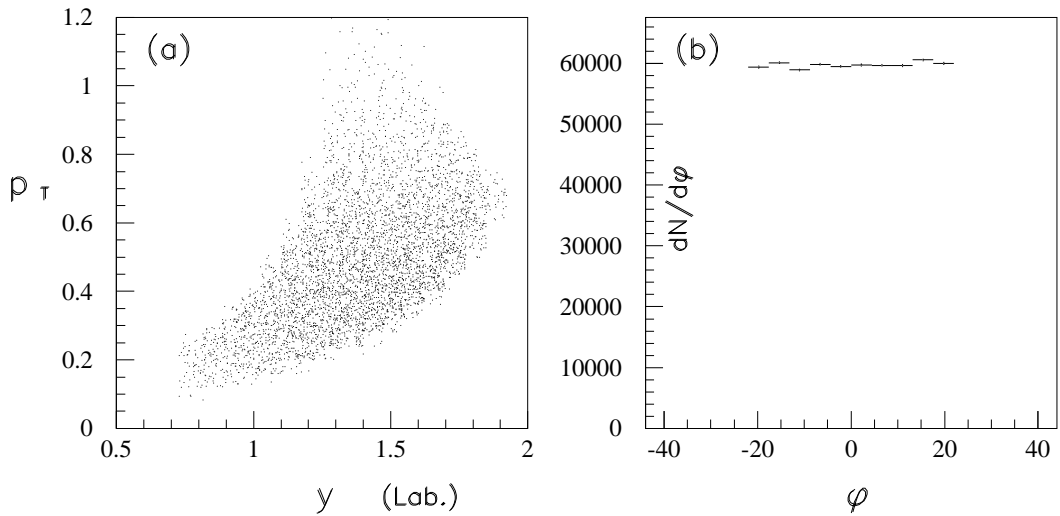}
\end{center}
\vskip -9.2cm
\caption{{\bf Phase space generated by the CERES code, with the simplified 
cuts shown in eq. (\ref{a2exp}) and eq.(\ref{a1exp}).}}
\label{ptxyphi}
\end{figure}
\vskip .7cm
The angles are measured in degrees and the momenta in GeV/c. The single
inclusive distribution  cuts are specified by 

\begin{equation}
A_1({\boldmath k})= \Theta(14 <\theta_{lab}< 28) 
\Theta(p_{lab}< 2.9\;{\rm GeV/c})  \Theta(y_{min}> 0.75) 
\; \; .\label{a1exp}\end{equation} 

	It can be seen from Figure 1 that the phase-space generated with 
the above cuts reproduces very closely that covered by the 
experiment \cite{vince}. Only 
a few excess of generated particles can be seen at low transverse 
momentum $p_T$ (or $k_{T_i}$).

To assess  the statistical significance of
the differences between the fits obtained assuming
resonance and non-resonance dynamics, we computed 
the $\chi^2$ goodness of fit, estimating this variable as 
previously\cite{pgplb95,zajc} 
\begin{equation} 
\chi^2(i,j) =\frac{ [A(i,j) - {\cal N_\chi}^{-1} C_{th}(i,j)  B(i,j)]^2 }
{ \{ [\Delta A(i,j)]^2 + [ {\cal N_\chi}^{-1} C_{th}(i,j)  
\Delta B(i,j)]^2 \} }
\; \; ,\label{chi2}\end{equation}
where  ${\cal N}_\chi$ is a normalization factor 
which is chosen to minimize the 
average $\chi^2$ and depends on the range in the
($q_T,q_L$) plane under analysis. The indices $i,j$ refer to the
 the corresponding $q_T,q_L$
bins, in each of which the experimental correlation function
is given by  
\begin{equation} 
C_{E}(i,j) = {\cal N_\chi}\frac{A(i,j)}{B(i,j)} \;  ; \; 
\Delta C_{E}(i,j) = C_{E}(i,j) \sqrt{ \left(
\frac{\Delta A(i,j)}{A(i,j)}\right)^2 + 
\left(\frac{\Delta B(i,j)}{B(i,j)}\right)^2 } 
\; \; .\label{cexp}\end{equation}
The numerator $A(i,j)\pm \Delta A(i,j)$ and denominator
$B(i,j)\pm \Delta B(i,j)$ in Eq. (\ref{chi2}) and (\ref{cexp}) 
were obtained from V. Cianciolo\cite{vince,vince:priv}, understanding that
the data in this form are preliminary and subject to further
final  analysis. Use is made of its preliminary form mainly for testing the 
sharpness of the method. Note that in the present analysis we are not 
including the errors 
associated to the theoretical correlation function generated by the Monte Carlo 
importance sampling in CERES. All calculations, however, were performed by 
taking high statistics runs only, making it reasonable to 
consider those errors as negligible.

 Analogously to the procedure adopted 
in Ref. \cite{pgplb95}, minimization of the average $\chi^2$ was performed 
by exploring the parameter space of  $R_T$ and $\Delta \tau$ and computing 
the $\langle \chi^2 \rangle$, averaging over a grid of nearly 30x30 bins 
in the $(q_T,q_L)$ plane 
in the relative momentum region $0.005 < q_T,q_L < 0.605$ GeV/c, 
binned with  $\delta q_T=\delta q_L=0.02$ GeV/c. 
A very meticulous investigation was performed to find the most probable 
region where the minimum ($R_{T_0},\Delta\tau_0$) could be located. 

\begin{figure}[htb]
\epsfxsize=0.6\textwidth
\begin{center}
\leavevmode
\epsfbox{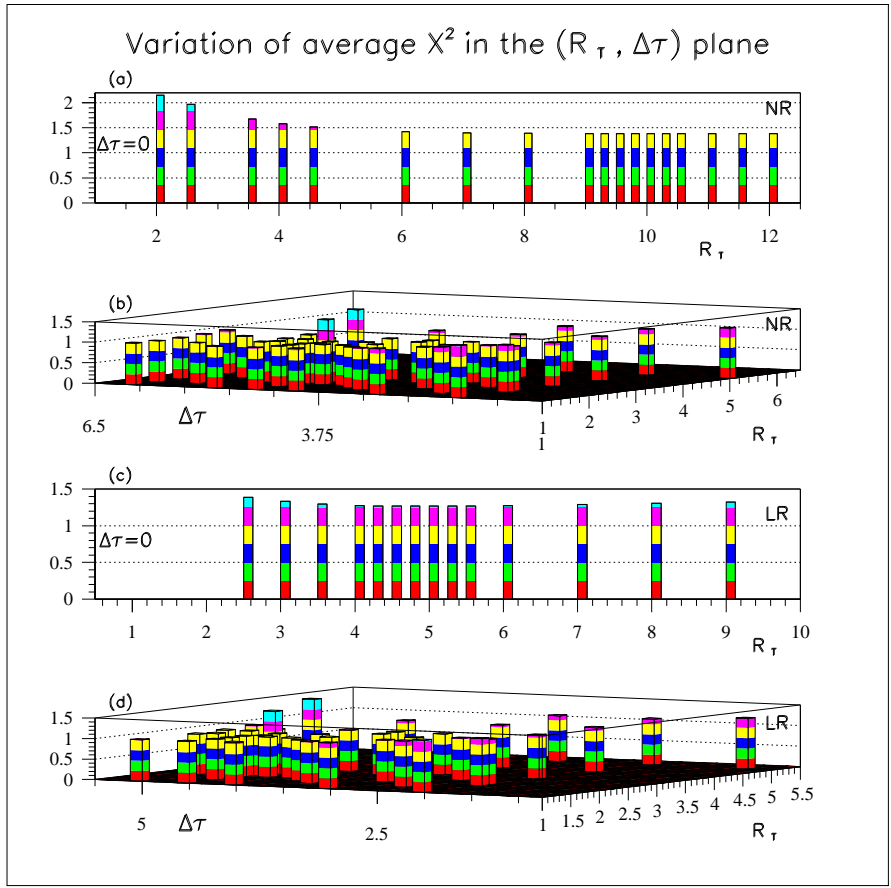}
\end{center}
\vskip -1.0cm
\caption{{\bf Zone in the ($R_{T},\Delta\tau$) plane investigated, 
leading to the determination of the most probable region where the 
minimum $\langle \chi^2 \rangle$, associated to ($R_{T},\Delta\tau$), 
could be located. Part (b) corresponds to the dynamics ignoring the 
contribution of $K^\star$ and part (d) to including 
their contribution to the kaon yield. Similarly, parts 
(a) and (c) correspond to non-resonance and resonance 
cases, respectively, but fixing $\Delta\tau=0$, and optimizing only 
$R_{T}$.  }}
\label{chi2rtau}
\end{figure}

\vskip 0.7cm

Figure 2 illustrates most of the investigated region in the 
($R_{T},\Delta\tau$) plane. Parts (a) and (c) will be discussed latter. 
In the vicinity of the minimum, the parameters of the 
quadratic surface
$
\langle \chi^2(R_T,\Delta\tau)\rangle =\chi^2_{min} +  \alpha (R_T-R_{T_0})^2
 + \beta (\Delta\tau-\Delta\tau_0)^2 
$ were determined.   
The results of 
such investigation are given in Table 1, where the radius 
parameters are measured in fm and time  intervals in fm/c. 

\vskip .9cm
\begin{center}
{\bf TABLE 1: 2D-$\chi^2$ Analysis of Kaon Decoupling Geometry} \\
\begin{small} 

\vskip 0.7cm

\begin{tabular}{|c|c|c|}
\hline 
$\chi^2(R_{T},\Delta \tau)$  & No Res. $(f_{K_{dir}}=1)$ & 
LUND Res. $(f_{K_{dir}}=f_{K/K^*}=0.5)$ \\
\hline
\multicolumn{3}{|c|}{Optimized $R_T$ and $\Delta \tau$}\\
\hline
$\langle \chi^{2}_{min} \rangle _{30 \times 30}$ & 1.03 & 1.02 \\
$\langle \chi^{2}_{min} \rangle _{10 \times 10}$ & 1.17 & 1.30 \\
$R_{T0}$  & 2.19$\pm$ 0.76 & 1.95 $\pm$ 0.89 \\
$\Delta \tau_0 $ & 4.4$\pm$ 2.0 & 4.4$\pm$ 2.6 \\
$\alpha$ & 0.0410 & 0.0299  \\
$\beta$ & 0.0058 & 0.0034 \\
\hline
\multicolumn{3}{|c|}{Optimized $R_T$ ($\Delta \tau=0)$}\\
\hline
$\langle \chi^{2}_{min} \rangle_{30 \times 30}$ & 1.29 & 1.33 \\
$\langle \chi^{2}_{min} \rangle_{10 \times 10}$ & 4.04 & 2.92 \\
$R_{0_T}$ & $\sim$ 10.6 & $\sim$ 4.8 \\
$\alpha$ & 0.0003 & 0.0280  \\
\hline
\end{tabular}

\end{small}
\end{center}
\bigskip

\vskip 0.5cm

	The errors appearing in Table 1 were estimated following the 
prescription of Ref.\cite{pgplb95}, which considered the $\chi^2$  
over $N$ bins as a random variable and, for large $N$, 
approximated the distribution of the mean $\chi^2$ per bin by 
$P(\chi^2)\propto \exp[-(\chi^2-1)^2/2\sigma^2]$, with 
rms width $\sigma=\sqrt{2/N} \approx 0.048$, for the $N=855$ grid under 
consideration (i.e., subtracting from the 
original 900 the empty bins and the number of degrees of freedom 
consumed in the $\chi^2$ analysis itself). 
Inserting the expression for $\langle \chi^2(R_T,\Delta\tau)\rangle$ 
in the above paraboloid into the asymptotic form of the
$\chi^2$ distribution for large $N$, the likelihood
for the parameter $R_T$ to have a value near the
minimum is approximately $\propto \exp[-\alpha^2(R_T-R_{T_0})^4/2\sigma^2]$. 
Therefore the estimated error on the radius is 
$\Delta R\approx \{\sqrt{2}[\Gamma(3/4)/\Gamma(1/4)] \sigma/\alpha\}^{1/2}
\approx 0.7( \sigma/\alpha)^{1/2}$, and similarly the error on the proper 
time interval is $0.7 (\sigma/\beta)^{1/2}$.

	Comparing Table 1 with Ref.\cite{pgplb95}, we may see that 
the optimized value of $\Delta\tau$, 
the decoupling time interval, is estimated to be about the same as in 
the pion case. However, the 
transverse size of the kaon emission region seems to be half that 
of the pions. This results agrees with the experimental fit 
to the data and, as was stated in Ref.\cite{vince}, it 
could be reinforcing the suggestion in Ref. \cite{nagaetal}, 
according to which kaons could decouple earlier than pions 
due their small cross section for interacting with nuclear 
matter. 

	We see from Table 1 that the optimization in both scenarios result 
in similar values for $\langle \chi^2 \rangle$ over  855 bins, 
although using the optimized parameters, we see smaller 
$\langle \chi^2 \rangle$ for the non-resonance
scenario in a smaller ($10 \times 10$) grid. 
Just to illustrate the similarities, we 
can see in Figure 3 the two-dimension correlation functions 
$C(q_T,q_L)$, corresponding to data and to the theoretical 
values generated with the optimized values shown in Table 1. 
\vskip -1.2cm
\begin{figure}[htb]
\epsfxsize=0.7\textwidth
\begin{center}
\leavevmode
\epsfbox{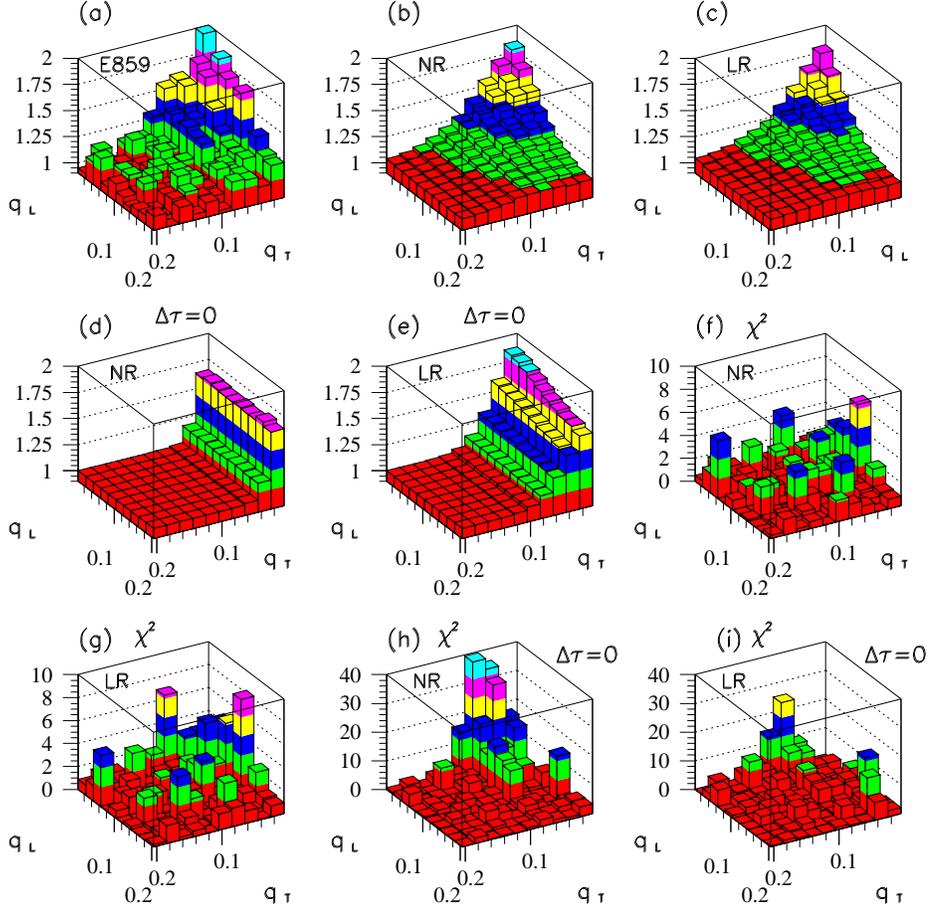}
\end{center}
\vskip -1.9cm
\caption{{\bf The preliminary 
E859 Gamow corrected data are shown in part (a). Part (b) shows 
theoretical correlation functions, $C(q_T,q_L)$, filtered with the E859 
acceptance for the case with no resonances (NR) and part (c) corresponds 
to the inclusion of Lund resonance (LR) fractions; 
the corresponding distribution of $\chi^2(q_T,q_L)$ are in (f) and (g), 
respectively. 
Similarly, when fixing  $\Delta\tau = 0$, results for 
the generated $C(q_T,q_L)$ are shown in parts (d) (NR) and (e) (LR), with 
$\chi^2(q_T,q_L)$ distribution in (h) and (i), respectively. }}
\label{2d}
\end{figure}

	From the above discussion, similarly to what happened in the 
pion case, we see that not enough separation is found, neither from 
the 2-D projection  
alone, nor by conjugating it to the average $\chi^2$ analysis. However, 
in Ref.\cite{pgplb95} it has already been recalled that a most direct measure 
of the goodness of fit could be achieved
by means of $n_\sigma=|\langle \chi^2_{min} \rangle - 1|/\sigma$, 
the number of standard deviations from unity
of the average $\chi^2$ per degree of freedom.
Since $n_\sigma$ depends 
on the range of $q$ under  analysis, we followed the steps of 
Ref. \cite{pgplb95} and studied its behavior by varying 
the range of the analysis to restricted $(q_T,q_L)$ domain, 
ranging from a  $2\times2$ grid, 
corresponding to $0.025<q_T,q_L<0.045$ GeV/c ,
to $3\times3$, $4\times4$, etc. as shown in Figure 4. 
For each $n\times n$ grid, $N=n^2$ is the number of 
degrees of freedom and the standard deviation is
expected to be $\sigma=\sqrt{2}/n$. 
The strong dependence of the number of standard deviations from unity
as a function of 
the range of the analysis is brought out clearly  in Figure 4. 

\vskip -1.9cm

\begin{figure}[htb]
\epsfxsize=0.9\textwidth
\begin{center}
\leavevmode
\epsfbox{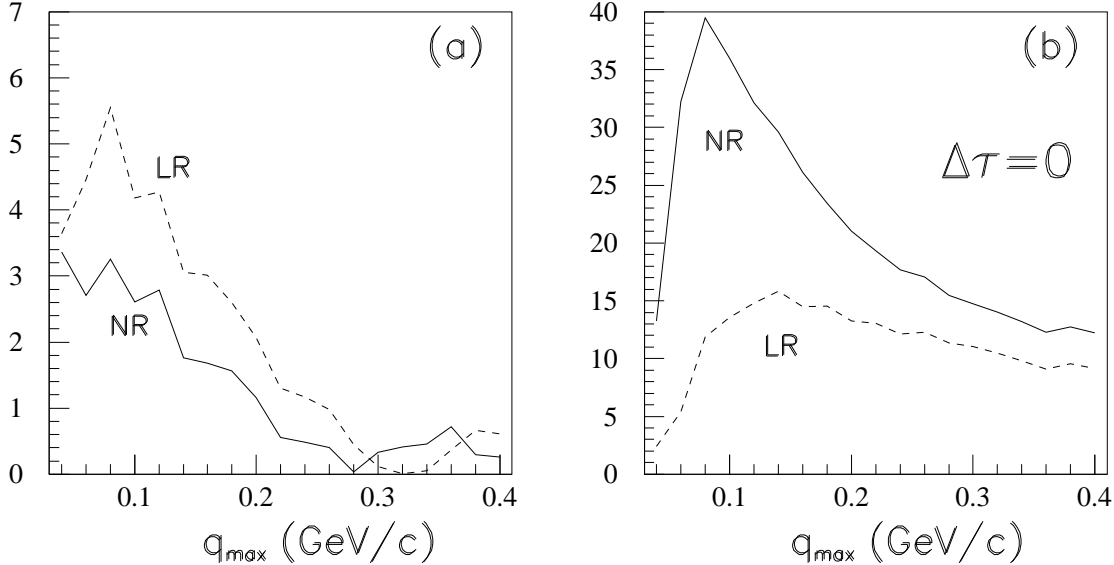}
\end{center}
\vskip -10.0cm
\caption{{\bf Number of standard deviations of $\langle \chi^2 \rangle$ from
unity for increasing number of bins. In part (a), $R_T$ and $\Delta\tau$ were 
optimized, whereas in (b), only $R_T$ was. }}
\label{dof}
\end{figure}

Although less striking than in the pion case, we see from Figure 4 
that it still is possible to separate the two scenarios, 
although none of them could be considered as a very good fit. This situation 
could become better in the near future with improving statistics, 
which could allow for smaller bin sizes. It is clear, however, 
that the non-resonance picture is closer to the preliminary data in 
all the range significant for interferometry, i.e., in the 
smaller domain $q_T,q_L<0.20$ GeV/c, where the correlation function
deviates significantly  from unity.
The two models yield similar fits in terms
of $\chi^2$ for $q_{max}>200$ MeV/c because in that large
domain both models trivially predict nearly unit correlation functions.

In Figure 4.(b) two curves signaled with $\Delta \tau = 0$ can be seen. 
This result corresponds to fixing the decoupling time interval 
to zero (instantaneous freeze-out) and searching 
for the optimized value of $R_T$. This test was performed
following the suggestion in Ref. \cite{vince}, according to which 
the preliminary 3-D experimental analysis in $q_0, q_T, q_L$ 
returned results for $R_T,R_L$ compatible with values obtained 
by the 2-D analysis, although the value of $\Delta \tau$ 
found could either be $\Delta \tau \approx R_T$ or $\Delta \tau=0$! 
We should recall 
that the experimental analysis had to project data in large bins 
(width of 180 MeV) \cite{vince}, in order to have enough 
statistics in the time direction which, by itself, 
would be responsible for dramatically weakening the interferometric 
signal. However, that ambiguous conclusion regarding the time interval 
was reached when a static Gaussian 
space-time parameterization was used in the experimental fit. 
We then decided to test what would be the response 
of the method to it, since we consider a different class of models, 
in which the longitudinal expansion is taken into account. The region 
of $R_T$ searched in its optimization, keeping $\Delta \tau = 0$, 
is shown in Figures 2 (a) and (c). The optimized values for $R_T$ 
are shown in Table 1. The errors, estimated with the aid of the 
asymptotic form of the $P(\chi^2)$  distribution, would not
apply to this case. The reason is that, for considering the average $\chi^2$ 
over N bins as a random variable with unit mean and rms width 
$\sigma=\sqrt{2/N}$, as discussed above, the $\chi(i,j)$ should be normal 
random variables with zero mean and unit rms width. We estimated this  
distribution in each case by running CERES for the optimized values. 
When fixing $\Delta\tau$  to be zero, however, the assumption 
made about the $\chi(i,j)$ distribution was not verified. For this 
reason, we prefer to simply show the optimized values of $R_T$ as 
approximate ones. The corresponding results can 
be seen both in Figures 3.(d) and (e) (with corresponding 
$\chi^2$ in (h) and (i)), as well as in Figure 4.b.  
From this last one, we see that our model  
completely excludes the instataneous emission. In particular, even 
in the region where no correlation is expected (roughly for 
$q_T, q_L >200$ MeV), the deviation with respect to data 
continues to be enormous. 

Furthermore, the above analysis nicely 
illustrates the important and well-known \cite{pg:nioc,hamapad,pratthpb} 
influence of the time spread in the {\it effective} transverse 
radius, $R_T$. Although its influence would be noticeable even for a static 
Gaussian parameterization of the space-time decoupling 
geometry\cite{pg:nioc}, models considering expanding systems strengthens 
the effect\cite{pg:nioc,hamapad}. 
In the present analysis, the time influence on $R_T$ can be inferred from  
the fact that the optimized radius 
increases considerably, trying to compensate for the strong constraint of 
zero emission time interval. 
For instance, when including the $K^\star$ contribution, its finite lifetime 
tries to circumvent the problem by introducing a 
non-zero time spread through the resonance decay, albeit the optimized 
$R_T$ is about twice the value without that constraint. 
This effect is, however, more dramatic in the non-resonance case, 
where no clear evidence about the location of the optimized value of $R_T$ 
can be seen from Figure 2.(a), since there 
is no way out to accommodate the instant emission constraint. 

We conclude that the multi-dimensional analysis proposed in 
Ref. \cite{pgplb95}, which has high resolving power in
the domain of physical interest in the case of pion interferometry, 
can still be applied to the case of kaon interferometry, 
although with less resolving power, due to the absence of contribution 
from long lived resonances. Finally, the above 
two-dimensional $\chi^2$ analysis indicates 
that, as far as the preliminary E859 data is concerned, 
expanding sources should be considered  at the AGS energy range, 
since expansion enhances the influence of the emission time interval on 
the transverse dimensions of the source and, from the present analysis, 
kaon sources emitting instantaneously are discarded. This should also be 
considered as an alert against the common practice of employing the static 
Gaussian parameterization to fit data since, by using it, the interpretation 
of the corresponding extracted parameters could be misleading or even wrong. 

\acknowledgments
We are very grateful to V. Cianciolo for
making his unpublished  data files and analysis available to us, 
and to W. Zajc and R. Soltz for many illustrative discussions. 
Helpful comments from M. Gyulassy 
and several discussions with P. Gouffon, M. C. 
Gonzalez-Garcia and R. Vazquez on practical matters are also 
thankfully acknowledged.

\end{document}